\documentclass[journal]{IEEEtran}
\usepackage{amsmath,amssymb}
\usepackage[dvips]{graphicx}
\usepackage{amsfonts}
\usepackage[mathscr]{eucal}
\usepackage{latexsym}
\usepackage{amsthm}
\usepackage{exscale}
\usepackage[mathscr]{eucal}
\usepackage{bm}
\usepackage[dvipsnames]{color}
\usepackage{cases}
\usepackage{epsfig}
\usepackage[center,small]{caption}
\usepackage{algorithm}
\usepackage{algorithmic}
\usepackage[verbose,nospace,sort]{cite}
\usepackage{tabularx}
\usepackage{multirow}
\usepackage{multicol}
\usepackage{balance}
\usepackage{bookmark}
\usepackage{amsmath}
\usepackage{amssymb}
\usepackage{mathrsfs}
\graphicspath{{./figs/}}

\usepackage{graphicx}

\usepackage[labelformat=brace]{subcaption}

\ifCLASSINFOpdf

\else

\fi

\scrollmode

\begin{document}

\title{Receding Horizon Optimization for Energy-Efficient UAV Communication}

\author{Jingwei~Zhang,
       Yong~Zeng,~\IEEEmembership{Member,~IEEE,}
        and~Rui~Zhang,~\IEEEmembership{Fellow,~IEEE}
\vspace{-0.8cm}

\thanks{
J. Zhang and R. Zhang are with the Department of Electrical and
Computer Engineering, National University of Singapore, Singapore 117583 (e-mail:~jingwei.zhang@u.nus.edu, elezhang@nus.edu.sg).}
\thanks{Y. Zeng is with the National Mobile Communications Research Laboratory, School of Information Science and Engineering, Southeast University, China (e-mail:~yong$\_$zeng@seu.edu.cn).}
}

\maketitle

\begin{abstract}
In this letter, we study a wireless communication system with a fixed-wing unmanned aerial vehicle (UAV) employed to collect information from a group of ground nodes (GNs). Our objective is to maximize the UAV's energy efficiency (EE), which is defined as the achievable rate among all GNs per unit propulsion energy consumption of the UAV. To efficiently solve this problem with continuous-time functions, we propose a new method based on receding horizon optimization (RHO), which significantly reduces the computational complexity compared to the conventional time discretization method. Specifically, we sequentially solve the EE maximization problem over a moving time-window of finite duration, for each of which the number of optimization variables is greatly reduced. Simulation results are provided to show the effectiveness of the proposed method.
\end{abstract}
\begin{IEEEkeywords}
UAV communication, energy efficiency, time discretization, trajectory optimization, receding horizon optimization.
\end{IEEEkeywords}

\section{Introduction}

Unmanned aerial vehicle (UAV)-enabled communication platforms are envisioned as a promising candidate for future wireless systems, thanks to their on-demand and swift deployment, great flexibility, high mobility and line-of-sight (LoS) dominant communication links with the ground nodes (GNs) \cite{zeng2016wireless}, \cite{zeng2019accessing}. To realize the full potential of UAVs for wireless communications, prior research can be generally classified into two categories, namely, the placement optimization for quasi-stationary UAVs and the trajectory design for flying UAVs \cite{6863654,7572068,7932157,8247211}.

In particular, the optimization of UAV trajectory offers an important new design degree of freedom (DoF) for communication performance enhancement, and has opened a new research paradigm into UAV trajectory and communication co-design. However, such co-design problems generally involve continuous-time variables and are highly non-convex, thus rendering them difficult to solve. To tackle this new type of optimization problems, various techniques have been proposed, such as trajectory discretization, block coordinate descent (BCD), and successive convex approximation (SCA) \cite{zeng2019accessing}, \cite{7932157,8247211,zeng2017energy,zeng2018energy}. Among them, trajectory discretization aims to convert the continuous-time trajectory variables approximately into discrete counterparts, so as to reduce the problem size to be finite. There are mainly two approaches for trajectory discretization, namely, \emph{time discretization} \cite{zeng2017energy} and \emph{path discretization} \cite{zeng2018energy}. Specifically, with time discretization, the UAV flight duration is equally divided into a finite number of time slots that are sufficiently small, so that the UAV trajectory can be approximated by a sequence of line segments with ordered waypoints, each specifying the start/end UAV location over two consecutive time slots. In contrast, with path discretization, the UAV path is discretized into consecutive line segments that are in general of different durations. Moreover, to ensure a high discretization accuracy, the length of each line segment in both discretization methods should be no greater than a certain threshold, such that the distance between the UAV and GNs can be regarded to be approximately constant within each line segment.

However, one critical issue of the above trajectory discretization methods lies in the large number of line segments and corresponding design variables (such as per-segment waypoints and time durations) \cite{8443133,you20193d,8648498}, which increase drastically with the UAV flight duration/distance. Furthermore, these variables are generally coupled with each other due to the various UAV mobility constraints, and thus cannot be optimized separately or in parallel. Hence, effective solution to reduce the computational complexity for designing communication-oriented UAV trajectory is crucial. In \cite{shen2018multi}, the authors proposed the alternating direction method of multipliers (ADMM) to decompose the UAV trajectory optimization problem involving $M$ UAVs into $M$ parallel problems, each for one UAV. Note that the existing works for complexity reduction mainly focus on problem decomposition across UAVs, which cannot be applied to resolve the fundamental complexity bottleneck due to the large number of variables even for individual UAV's trajectory optimization.

Motivated by the above, we study in this letter a wireless communication system enabled by a fixed-wing UAV, which is employed to collect data from a group of distributed GNs (e.g., sensors). Our objective is to maximize the UAV's energy efficiency (EE), which is defined as the ratio of the achievable communication throughput among all GNs over the (propulsion) energy consumption of the UAV. To reduce the computational complexity arising from the large number of optimization variables, we propose a new method based on the technique of receding horizon optimization (RHO) \cite{1657504,bellingham2002receding,grancharova2015uavs,camacho2013model}. The key idea of RHO is to use a moving time-window of finite duration to successively solve the EE maximization problem. Within each time-window, different from the conventional time discretization method where the maximum segment length is set to be identical, two different values for the maximum segment length are used in the RHO method to reduce the number of optimization variables, thus leading to a lower computational complexity.

It is worth mentioning that RHO has been applied for UAV control and navigation design \cite{bellingham2002receding}, \cite{grancharova2015uavs}, but without being fully exploited to enhance the performance of UAV-enabled communication system. To our best knowledge, the current work is the first that applies RHO for joint UAV trajectory and resource allocation optimization, which is applicable to UAV communications in general.

\section{System Model and Problem Formulation}

We consider a UAV-enabled wireless communication system, where a fixed-wing UAV is employed as a flying access point to collect information from $L$ GNs, which are denoted by the set $\mathcal{L}=\{1,\cdots,L\}$. Let the horizontal location of GN $l \in \mathcal{L}$ be denoted as $\textbf{w}_l \in \mathbb{R}^{2 \times 1}$. The UAV is assumed to fly at a given altitude $H$ and follow a periodic trajectory with period $T$, which is denoted as $\textbf{q}(t)\in \mathbb{R}^{2 \times 1}$, $0 \leq t \leq T$. Thus the time-varying distance between the UAV and GN $l$ is $d_l(t)=\sqrt{H^2+ ||\textbf{q}(t)-\textbf{w}_l||^2}$.

Let $h_l(t)$ denote the channel coefficient from GN $l$ to the UAV at time $t$, which can be decomposed as
\begin{align}
h_l(t)=\sqrt{\beta_l(t)} \tilde{h}_l(t),
\end{align}
where $\beta_l(t)$ represents the large-scale channel power accounting for the distance-dependent path loss, and $\tilde{h}_l(t)$ is a random variable with $\mathbb{E}[|\tilde{h}_l(t)|^2]=1$ accounting for the small-scale fading due to signal reflection/scattering. Furthermore, the large-scale channel power $\beta_l(t)$ can be modelled as $\beta_l(t)=\beta_0 d_l^{-\tilde{\alpha}}(t)$, where $\beta_0$ is the path loss at a reference distance of $d_0=1$ m and $\tilde{\alpha}$ denotes the path loss exponent.

We assume that time-division multiple access (TDMA) is employed among different GNs to access the wireless channel. Define a binary variable $\lambda_l(t)$, which is 1 if GN $l$ is scheduled for transmission to the UAV at time $t$, and $0$ otherwise. We then have the following constraints
\begin{align}
\label{d01}
\sum_{l=1}^L \lambda_l(t) \leq 1, \ \ \forall t \in [0,T],~~~~~~~~\\
\label{d02}
\lambda_l(t) \in \{0,1\}, \ \ \forall t \in [0,T] , l \in \mathcal{L}.
\end{align}
By denoting the transmit power of GN $l$ as $P_l$, the bandwidth-normalized achievable rate between GN $l$ and UAV in bits per second (bps) can be expressed as
\begin{align}
\tilde{R}_l(t)= B \lambda_l(t) \log_2\left(1+\frac{P_l|h_l(t)|^2}{\sigma^2 \Gamma}\right),
\end{align}
where $\sigma^2$ is the noise power at the receiver, and $\Gamma>1$ accounts for the gap from the channel capacity due to the practical modulation and coding scheme employed, and $B$ denotes the available bandwidth. Therefore, the accumulated communication throughput for GN $l$ during one period $T$ is $\tilde{R}_l= \int_0^T \tilde{R}_l(t) dt$.

It is noted that $\tilde{R}_l$ is a random variable due to the randomness of $\tilde{h}_l(t)$, whose exact distribution is difficult to obtain. Therefore, we consider its expected value $\bar{R}_l=\mathbb{E}[\tilde{R}_l]$ for simplicity. While finding the exact expression of $\bar{R}_l$ is also challenging, the following holds \cite{zeng2018energy},
\begin{align}
\!\!\!\!\!\!\!\!\!\!\!\!\!\!\!\!\!\!\!\!\!\!\!\!\bar{R}_l(\{\lambda_l(t)\},\{\textbf{q}(t)\}  )
\stackrel{(a)}{\leq} \int_0^T B \lambda_l(t) \log_2\left(1+\frac{P_l\mathbb{E}[|h_l(t)|^2]}{\sigma^2 \Gamma} \right)dt \nonumber\!\!\!\!\!\!\!\!\!\!\!\!\!\!\!\!\!\!\!\!\\
\stackrel{(b)}{=}\int_0^T B\lambda_l(t) \log_2 \left(1+\frac{ \gamma_l}{(H^2+ ||\textbf{q}(t)-\textbf{w}_l||^2)^{\alpha}}\right) dt \nonumber\!\!\!\!\!\!\!\!\!\! \\
\triangleq  \int_0^T R_l(t)dt \triangleq R_l(\{\lambda_l(t)\},\{\textbf{q}(t)\} ),~~~~~~~~~~~~~\!
\end{align}
where $(a)$ follows from the Jensen's inequality, and $(b)$ holds due to the fact that $\mathbb{E}[|h_l(t)|^2]=\beta_l(t)$ and we have defined $\gamma_l \triangleq P_l \beta_0/ (\sigma^2 \Gamma)$ as the reference signal-to-noise ratio (SNR) at a reference distance of $d_0=1$ m, and $\alpha=\tilde{\alpha}/2$.

On the other hand, the energy consumption of the UAV constitutes two main components, namely, the propulsion energy to ensure that the UAV maintains aloft and support its mobility, as well as the communication-related energy for circuitry, signal processing, radiation, and so on. Since the communication-related power is much smaller than the propulsion power for practical UAVs, it is ignored for simplicity. Furthermore, the propulsion energy consumption of fixed-wing UAVs can be modelled as \cite{zeng2017energy}
\begin{align}
\!\!\!\!\!\!\!\!\!\!\!\!\!\!\!\!\!\!\!\!\!\!\!\!\!\!\!\!E(\{\textbf{v}(t)\}, \{\textbf{a}(t)\} )\approx
\int_0^T \bigg(c_1 ||\textbf{v}(t)||^3 +
\frac{c_2}{||\textbf{v}(t)||}\Big(1+\!\!\!\!\!\nonumber \\
\frac{||\textbf{a}(t)||^2}{g^2}\Big)  \bigg)dt
+\frac{1}{2} m \left( ||\textbf{v}(T)||^2-||\textbf{v}(0)||^2 \right),\!\!\!\!\!\!\!\!\!\!\!\!\!\!\!
\end{align}
where $\textbf{v}(t)=\dot{\textbf{q}}(t)$ and $\textbf{a}(t)=\ddot{\textbf{q}}(t)$ denote the UAV's velocity and acceleration vectors, respectively, $c_1$ and $c_2$ are two parameters related to the aircraft design, $g$ is the gravitational acceleration with nominal value of 9.8 m/s$^2$, and $m$ is the total mass of the UAV.

Our objective is to maximize the UAV's EE, which is defined as the common communication throughput achievable by all GNs normalized by the UAV's propulsion energy consumption. Mathematically, we define the EE as
\begin{align}
EE(\{\lambda_l(t)\},\{\textbf q(t)\},\{\textbf{v}(t)\}, \{\textbf{a}(t)\})  \triangleq~~~~~~~~~~~~~~~~~~~  \nonumber \\
 \frac{ {\min_{l\in \mathcal L} R_l(\{\lambda_l(t)\},\{\textbf q(t)\} )}}{{E(\{\textbf{v}(t)\}, \{\textbf{a}(t)\} )}}.~~~~~
\end{align}
The optimization problem is then formulated as
\begin{subequations}
\begin{align}
\!\!\!\!\!\mathrm{(P1)} \max_{\substack{ \{\lambda_l(t)\}, \{\textbf{q}(t)\}, \\\{\textbf{v}(t)\}, \{\textbf{a}(t)\} }}~~EE(\{\lambda_l(t)\},\{\textbf q(t)\},\{\textbf{v}(t)\}, \{\textbf{a}(t)\}) \nonumber\\
\label{p1001}
\mathrm{s.t.}~~\sum_{l=1}^L \lambda_l(t) \leq 1, \ \ \forall t \in [0,T],~~~~~~~~~~~\!~~~~~~~~ \\
\label{p1002}
\lambda_l(t) \in \{0,1\}, \ \ \forall t \in [0,T] , l \in \mathcal{L},~~~\!~~~\!~~~~~\\
\label{p1003}
\textbf{v}(t)=\dot{\textbf{q}}(t),~\textbf{a}(t)=\ddot{\textbf{q}}(t), \ \ \forall t \in [0,T],~~~~~\\
\label{p1004}
V_{\mathrm{min}} \leq ||\textbf{v}(t)|| \leq V_{\mathrm{max}}, \ \ \forall t \in [0,T],~~~~~~~ \\
\label{p1005}
||\textbf{a}(t)|| \leq a_{\mathrm{max}}, \ \ \forall t \in [0,T], ~~~~~~\!~~~~~~~~~~~\\
\label{p1006}
\textbf{v}(0)=\textbf{v}(T), \ \
\textbf{q}(0)=\textbf{q}(T),~~~~~~~~~~~~\!~~~~~
\end{align}
\end{subequations}
where $V_{\mathrm{min}}$, $V_{\mathrm{max}}$, and $a_{\mathrm{max}}$ denote the UAV's minimum speed, maximum speed, and maximum acceleration, respectively. The constraints in \eqref{p1006} ensure that the UAV returns to its initial location and velocity after each period of duration $T$. Problem (P1) is challenging to solve for two main reasons. Firstly, it involves continuous functions over $t$, which essentially involve an infinite number of variables. Secondly, the problem is non-convex with respect to the trajectory and time allocation due to the non-concave objective function and the non-convex constraints \eqref{p1002}-\eqref{p1004}. In the following, we first give a brief overview of the existing approach to address this problem based on time discretization and SCA applied over the entire time horizon $T$, which requires high computational complexity even for moderately large $T$. Then we propose a new method based on RHO to significantly reduce the complexity.

\vspace{-0.2cm}
\section{Conventional Method to solve (P1)}
\label{conventional}

To convert (P1) into a more tractable form with a finite number of variables, \cite{zeng2017energy} proposed a time discretization technique, where the entire time horizon $[0,T]$ is divided into $N$ time slots, each with equal length $\delta_1=T/N$. Let $t_n= n \delta_1$, $n=1,\cdots,N$. The continuous UAV trajectory $\{\textbf{q}(t)\}$ is then approximated by a finite number of line segments with waypoints $\textbf{q}[n]\triangleq\textbf{q}(t_n)$, together with the velocity and acceleration vectors $\textbf{v}[n]\triangleq\textbf{v}(t_n)$, $\textbf{a}[n]\triangleq\textbf{a}(t_n)$. Note that to ensure sufficient accuracy, each line segment should not exceed a certain length $\Delta_1$, which is chosen to be sufficiently small such that the distances between the UAV and GNs are approximately unchanged within each line segment or time slot. Thus, for any given $\Delta_1$, the time slot length $\delta_1$ should be chosen to cater for the worst-case scenario, i.e, when the UAV files with the maximum speed $V_{\mathrm{max}}$, such that $\delta_1 V_{\mathrm{max}} \leq \Delta_1$. As a result, the minimum number of time slots required is $N=\lceil T V_{\mathrm{max}}/\Delta_1 \rceil$, where $\lceil b \rceil$ denotes the minimum integer no smaller than $b$. Moreover, with such a time discretization approach, the binary constraints in \eqref{p1002} can be readily circumvented by introducing the variable $\rho_l[n]$, which denotes the fraction of the time allocated to communicate with GN $l$ at time slot $n$. Specifically, it is not difficult to see that it is always possible to find $\rho_l[n]$ satisfying \eqref{p1001} and \eqref{p1002} as long as the following constraints are met
\begin{align}
\sum_{l=1}^L \rho_l[n] \leq 1, \ \ \forall n, ~~~~~~~~~~~~~~~~~\!\\
\rho_l[n] \geq  0, \ \ \forall n, l \in \mathcal{L}.~~~~~~~~~~~~~~
\end{align}
Furthermore, the instantaneous achievable rate can be approximated as
\begin{align}
R_l[n]=B\rho_l[n] \log_2 \left(1+\frac{ \gamma_l}{(H^2+||\textbf{q}[n]-\textbf{w}_l||^2)^{{\alpha}}} \right).
\end{align}
In addition, the linear-state space approximation of the trajectory characterization in \eqref{p1003} is given by \cite{zeng2017energy}
\begin{align}
\!\!\!\!\!\!\!\!\!\textbf{v}[n+1]=\textbf{v}[n]+\textbf{a}[n] \delta_1, \ \ n=1,\cdots,N-1,~~~~~~\\
\!\!\!\!\!\!\!\textbf{q}[n+1]=\textbf{q}[n]+\textbf{v}[n]\delta_1+\frac{1}{2} \textbf{a}[n] \delta_1^2, n=1,\cdots,N-1.\!\!\!\!\!\!\!\!\!\!
\end{align}
After the above discretization, although (P1) is still non-convex due to the non-concave objective function and non-convex constraints \eqref{p1004}, it can be approximately solved by solving a sequence of convex optimization~problems~with monotonic convergence based on BCD and SCA techniques \cite{zeng2019accessing}, \cite{zeng2017energy}, \cite{8443133}. Specifically, with BCD, the UAV trajectory $\{\textbf{q}[n]\}$, $\{\textbf{v}[n]\}$, $\{\textbf{a}[n]\}$ and time allocation $\{\rho_l[n]\}$ are alternately updated with the other block fixed. For the sub-problem of trajectory update, SCA is adopted to convert it to a convex optimization problem. An overview of BCD and SCA techniques employed is given in \cite{zeng2019accessing} and the details are omitted for brevity.



\vspace{-0.2cm}
\section{Proposed Method Based on RHO}

Note that the discretization form of (P1) involves $(6+L)N+1$ optimization variables, while the complexity for solving (P1) via the conventional method is mainly due to solving the trajectory optimization sub-problem, which can be shown of complexity $O((TV_{\mathrm{max}}/\Delta_1)^{3.5})$ \cite{8443133}.

Although the complexity is polynomial of $T$ or $N=$ $\lceil TV_{\mathrm{max}}/\Delta_1 \rceil$, it is still practically exorbitant for even moderate values of $T$. For example, for $T$$=$$500$ s, $V_{\mathrm{max}}$$=$$30$ m/s, and $\Delta_1=10$ m, we then have $N=1500$, which leads to a high complexity of $O(1500^{3.5})\approx O(1.3\times 10^{11})$. To reduce the complexity for solving (P1) by the conventional method, we propose a new method based on RHO.

\begin{figure}
\vspace{-0.3cm}
\includegraphics[width=6in]{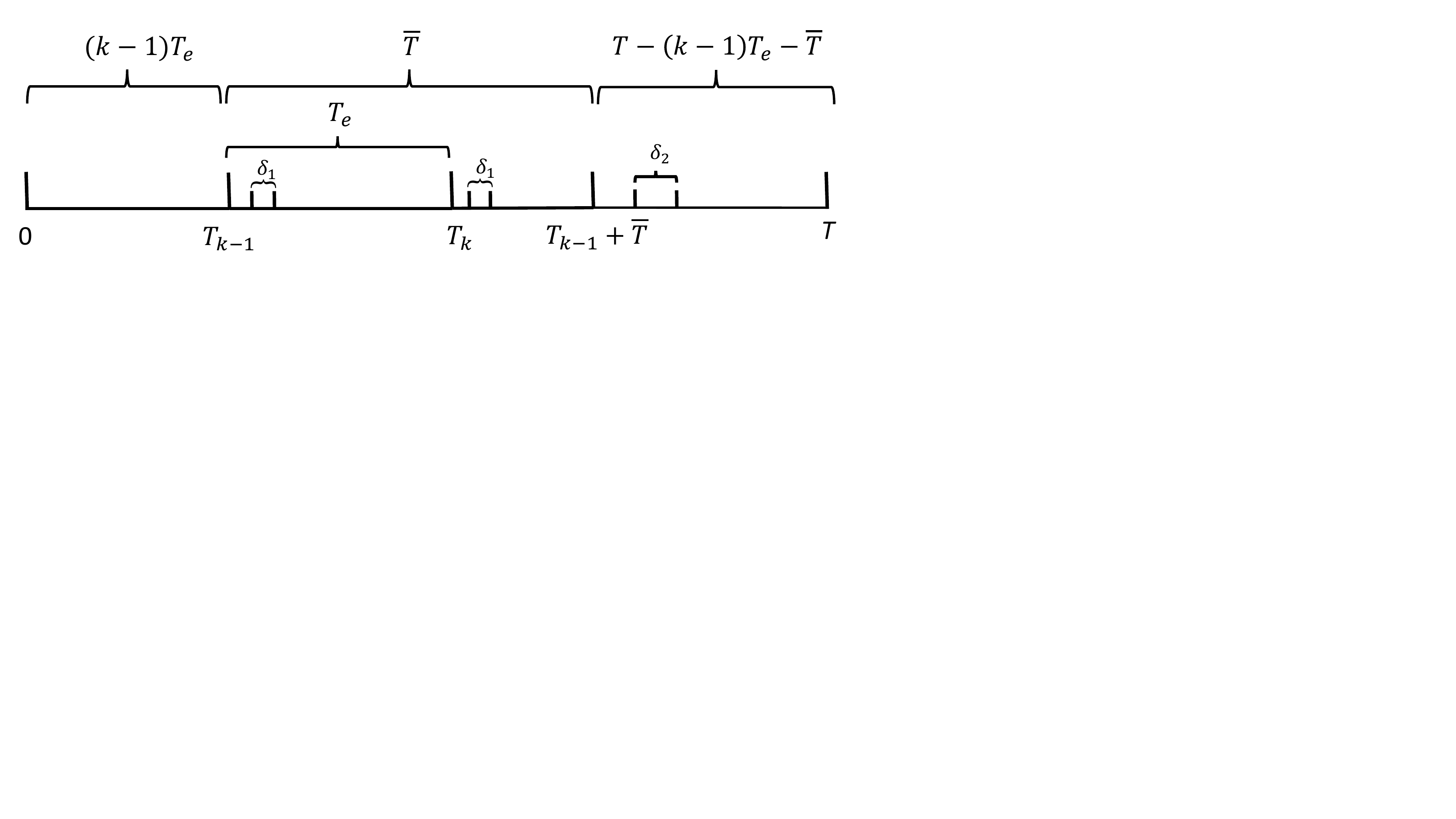}
\vspace{-6.5cm}
\caption{\label{rhc_syst}An illustration of the RHO-based method at the $k$th time-window. }
\vspace{-0.8cm}
\end{figure}

The main idea of RHO is that, instead of directly optimizing over the entire time horizon $[0, T]$, which results in a large number of time slots, (P1) can be iteratively optimized more efficiently over a moving time-window of duration $\overline{T} \ll T $ \cite{1657504}. Specifically, at each iteration, the entire time horizon $[0, T]$ is partitioned into three horizons, as illustrated in Fig. \ref{rhc_syst}. In the first horizon, trajectory and time allocation have already been obtained in the previous iterations or time-windows. While in the second horizon, trajectory and time allocation are to be optimized in the current iteration/time-window, where the trajectory is discretized with sufficiently high accuracy, i.e., with the maximum segment length set to $\Delta_1$. Last, the third horizon corresponds to the future time horizon, which is also optimized in the current time-window but with a much coarser discretization, e.g., with a larger maximum line segment length $\Delta_2>\Delta_1$. For convenience, we assume that $\Delta_2$ is an integer multiple of $\Delta_1$, i.e., $\Delta_2 = N_{\Delta} \Delta_1$ for certain integer $N_{\Delta}$. After solving the problem in each time-window, only the solution corresponding to the first $T_e \leq \overline{T}$ portion in the second horizon is actually executed by the UAV. Thus, the total number of time-windows is $K\triangleq \lceil(T-\overline{T})/T_e \rceil+1$. The details of formulating and solving the optimization problem in each time-window are given in the following.

Without loss of generality, we focus on the problem for the $k$th time-window, $k=1$$,$$\cdots$$,$$K$, to optimize the time allocation and trajectory from time $t=(k-1)T_e\triangleq T_{k-1}$ to $t=$$T_{k-1}$$+$$\overline{T}$.

Let the accumulated communication throughput for GN $l$ and the UAV energy consumption corresponding to the first horizon in Fig. \ref{rhc_syst}, i.e., from time $t=0$ to $t=T_{k-1}$, be denoted as $C_{l,{k-1}}$ and $E_{k-1}$, respectively. The second horizon from $t=T_{k-1}$ to $t=T_{k-1}+\overline{T}$ and the third from $t=T_{k-1}+\overline{T}$ to $T$ are discretized with time slot length $\delta_1\triangleq \Delta_1/V_{\mathrm{max}}$ and $\delta_2\triangleq \Delta_2/V_{\mathrm{max}}$, respectively, such that the number of time slots are $N_1=\lceil \overline{T}/\delta_1\rceil$ and $N_2=\lceil (T-T_{k-1}-\overline{T})/\delta_2\rceil$, respectively. Therefore, the time allocation and UAV trajectory can be re-discretized as $\rho_l[n]$$\triangleq$$\rho_l(t_n)$, $\textbf{q}[n]$$\triangleq$$\textbf{q}(t_n)$, $\textbf{v}[n]$$\triangleq$$\textbf{v}(t_n)$, and $\textbf{a}[n]$$\triangleq$$\textbf{a}(t_n)$, respectively, where for the second horizon, $t_n$$=$$T_{k-1}$$+$$n\delta_1$, $n$$=$$1,$$\cdots$$,$$N_1$, and for the third horizon, $t_n$$=$$T_{k-1}$$+$$\overline{T}+$$(n-N_1)\delta_2$, $n$$=$$N_1+1$$,$$\cdots$$,$$N_1+N_2$. As a result, the energy consumption of the UAV is approximated~as
\begin{align}
\!\!\!\!\!\!\!\!\!\!\!\!\!\!\!\!\!\!\!\!\!\!\!\!\!\!\tilde{E}=E_{k-1}+\delta_1 \sum_{n_1=1}^{N_1}\left(c_1 ||\textbf{v}[n]||^3 +\frac{c_2}{||\textbf{v}[n]||}\left(1+\frac{||\textbf{a}[n]||^2}{g^2} \right) \right)
\nonumber \!\!\!\!\!\!\!\!\!\!\!\!\!\!\!\!\!\!\!\!\!\!\!\!\!\!\!\!\!\! \\
+E_r,~~~~~~~~~~~~~~~~~~~~~~~~~~~~~~~~~~~~~~
\end{align}
where $E_r$ is the approximated energy consumption of the UAV from $t=T_{k-1} +\overline{T}$ to $t=T$ with given $\delta_2$. Then the problem for the $k$th time-window can be formulated as
\begin{subequations}
\begin{align}
\mathrm{(P-}k)\max_{\substack{\eta,\{\rho_l[n]\}, \{\textbf{q}[n]\}, \\\{\textbf{v}[n]\}, \{\textbf{a}[n]\}     } } ~~\frac{\eta}{\tilde{E}} \nonumber~~~~~~~~~~~~~\\
\label{pk001}
\!\!\!\!\!\!\!\!\!\!\!\!\!\!\mathrm{s.t.}~C_{l,k-1}+\delta_1 \sum_{n=1}^{N_1}R_l[n]+ \delta_2 \sum_{n=N_1+1}^{N_1+N_2}R_l[n] \geq \eta, \ \ l \in \mathcal{L},\!\!\!\!\! \\
\label{pk002}
\sum_{l=1}^L \rho_l[n] \leq 1, \ \ n=1,\cdots, N_1+N_2, ~~~~~~~~~~~~~~~~\\
\label{pk003}
\rho_l[n] \geq 0, \ \  n=1,\cdots, N_1+N_2, l \in \mathcal{L},~~~~~~~~~~~~~\\
\label{pk004}
\textbf{v}[n+1]=\textbf{v}[n]+\textbf{a}[n]\delta_1, \ \ n=1,\cdots,N_1-1,~~\!~~~\\
\label{pk005}
\!\!\!  \textbf{v}[n+1]=\textbf{v}[n]+\textbf{a}[n]\delta_2, \ \ n=N_1,\cdots,N_1+N_2-1,\!\!\! \\
 \label{pk006}
\!\!\!\!\!\!\!\!\!\!\!\!\!\!\! \textbf{q}[n+1]=\textbf{q}[n]+\textbf{v}[n]\delta_1+\frac{1}{2}\textbf{a}[n]\delta_1^2,\ \ n=1,\cdots,N_1-1, \!\!\!\!\!\\
\label{pk007}
 \textbf{q}[n+1]=\textbf{q}[n]+\textbf{v}[n]\delta_2+\frac{1}{2}\textbf{a}[n]\delta_2^2, \nonumber~~~~~~~~~~~~~~~~ \\ n=N_1,\cdots,N_1+N_2-1, \\
\label{pk008}
V_{\mathrm{min}}\leq ||\textbf{v}[n]|| \leq V_{\mathrm{max}}, \ \ n=1,\cdots,N_1+N_2,~~~ \\
\label{pk009}
||\textbf{a}[n]|| \leq a_{\mathrm{max}}, \ \ n=1,\cdots,N_1+N_2,~~~~~~~~~~~~~\\
\label{pk0010}
\textbf{q}[1]=\textbf{q}'_0, ~~~~~~~~~~~~~~~~~~~~~~~~~~~~~~~~~~~~~~~~~~~~~~~\!\\
\label{pk0011}
\textbf{v}[N_1+N_2]=\textbf{v}'[1],\ \
\textbf{q}[N_1+N_2]=\textbf{q}'[1],~~~~~~~~\!~
\end{align}
\end{subequations}
where $\eta$ represents the achievable rate by all GNs. Constraints \eqref{pk002}-\eqref{pk009} correspond to \eqref{p1001}-\eqref{p1005}. Moreover, $\textbf{q}'_0$ corresponds to the end UAV location $\textbf{q}[N_e+1]$ obtained after solving the previous $(k$$-1)$ problems to~ensure that~the obtained segments are connected, where $N_e\triangleq T_e/\delta_1$. Similarly, $\textbf{v}'[1]$ and $\textbf{q}'[1]$ in constraints \eqref{pk0011}, corresponding to \eqref{p1006}, are solutions to the problem of the first time-window to ensure that the UAV is able to return to the initial location at the end of the period $T$. Note that the total number of optimization variables for problem (P--$k$) is $(6$$+$$L)$$(N_1$$+$$N_2)$$+1$. All problems in the RHO-based method can be similarly solved by employing BCD and SCA techniques, for which the details are omitted for~brevity.
\vspace{-0.08cm}

Note that for each problem in the RHO-based method, the obtained trajectory and time allocation related to the third horizon from time $t=T_{k-1}+\overline{T}$ to $t=T$ are very coarse approximations due to the relatively large value of $\Delta_2$. Thus, after solving each problem, only the first $N_e$ time slots in $\{\textbf{q}[n]\}$, $\{\textbf{v}[n]\}$, $\{\textbf{a}[n]\}$, and $\{\rho_l[n]\}$
are actually executed by the UAV. The reason to include the third horizon in each problem is to provide a rough estimation of the trajectory and time allocation in the future to ensure that the UAV returns to the initial location and the target throughput of each GN is satisfied, while the accuracy of these coarse approximations will be further refined in subsequent~time-windows.
\vspace{-0.08cm}

Since all problems in the RHO-based method are solved based on BCD and SCA techniques, an initial trajectory is required for solving the problem in each time-window. For the first problem corresponding to $k=1$, since fixed-wing UAV is unable to stay stationary, the well-known Traveling Salesman Problem (TSP)-based trajectory initialization given in \cite{8443133} and \cite{8255824} is not applicable here. Therefore, the circular initial trajectory proposed in \cite{8247211} is employed with radius $r$ and speed $V=2 \pi r /T$. On the other hand, for problems in time-windows $k \geq 2$, the initial trajectory is set as the trajectory obtained after solving the previous $k-1$ problems.

By employing the RHO-based method, an efficient sub-optimal solution to problem (P1) can be obtained by sequentially solving reduced-size problems in different time-windows, and the key steps are summarized in Algorithm \ref{algo}.
\vspace{-0.25cm}
\begin{algorithm}
\caption{RHO-based method for solving (P1). \label{algo}}
1:~~Initialize $C_{l,0}=0$, $E_{0}=0$, and a circular trajectory. \\
2:~~\textbf{for} $k=1:K$ \\
3:~~~~Let $N_1=\lceil\overline{T}V_{\mathrm{max}}/\Delta_1\rceil$ and $N_2=\lceil (T-(k-1)T_e-$\\
\hspace*{0.65cm} $\overline{T})V_{\mathrm{max}}/\Delta_2\rceil$.\\
4:~~~~Design the initial trajectory for the $k$th problem.\\
5:~~~~Solve problem (P--$k$) for the $k$th time-window. Store the \\
\hspace*{0.65cm} obtained $\{\rho_l[n]\}_{n=1}^{N_e}$, $\{\textbf{q}[n]\}_{n=1}^{N_e}$, $\{\textbf{v}[n]\}_{n=1}^{N_e}$, $\{\textbf{a}[n]\}_{n=1}^{N_e}$. \\
6:~~~~Update $C_{l,k}$, $E_{k}$, $\textbf{q}_0'=\textbf{q}[N_e+1]$, $\textbf{v}'[1]=\textbf{v}[N_1+N_2]$ \\
\hspace*{0.65cm} and $\textbf{q}'[1]=\textbf{q}[N_1+N_2]$. \\
7:~~\textbf{end}
\end{algorithm}
\vspace{-0.6cm}

The overall complexity of the RHO-based method can be obtained as $O\big(\sum_{k=1}^{K-1}  (\overline{T}/\delta_1+$$(T-$$(k$$-$$1)T_e$$-\overline{T})/\delta_2  )^{3.5}+$ $( (T$$-(K$$-$$1)T_e)/\delta_1)^{3.5} \big)$. As the highest complexity~comes from the first time-window, the overall complexity~of the RHO-based method must be less than $O \big( K(\overline{T}/\delta_1$$+(T-\overline{T})/\delta_2    )^{3.5}    \big)$, which can be further approximated as
\begin{align}
O\left(\frac{T}{T_e} \left(\frac{(T+N_{\Delta}\overline{T})V_{\mathrm{max}}}{N_{\Delta} \Delta_1} \right)^{3.5}  \right).
\end{align}

It is noted that the values of $T_e$, $\overline{T}$, $\Delta_1$, and $\Delta_2$ need to be carefully chosen such that the proposed RHO achieves a desired complexity-performance trade-off, as shown in the next section. Finally, for the special case when $T_e$$=$ $\overline{T}$$=T$, the proposed RHO method becomes the same as the conventional time discretization method described in Section \ref{conventional}.

\section{Simulation Results}

\begin{figure}[htbp]
\vspace{-0.15cm}
\begin{minipage}[t]{0.45\linewidth}
\vspace{-3.4cm}
    \hspace*{-1.9cm}\includegraphics[width=2\linewidth]{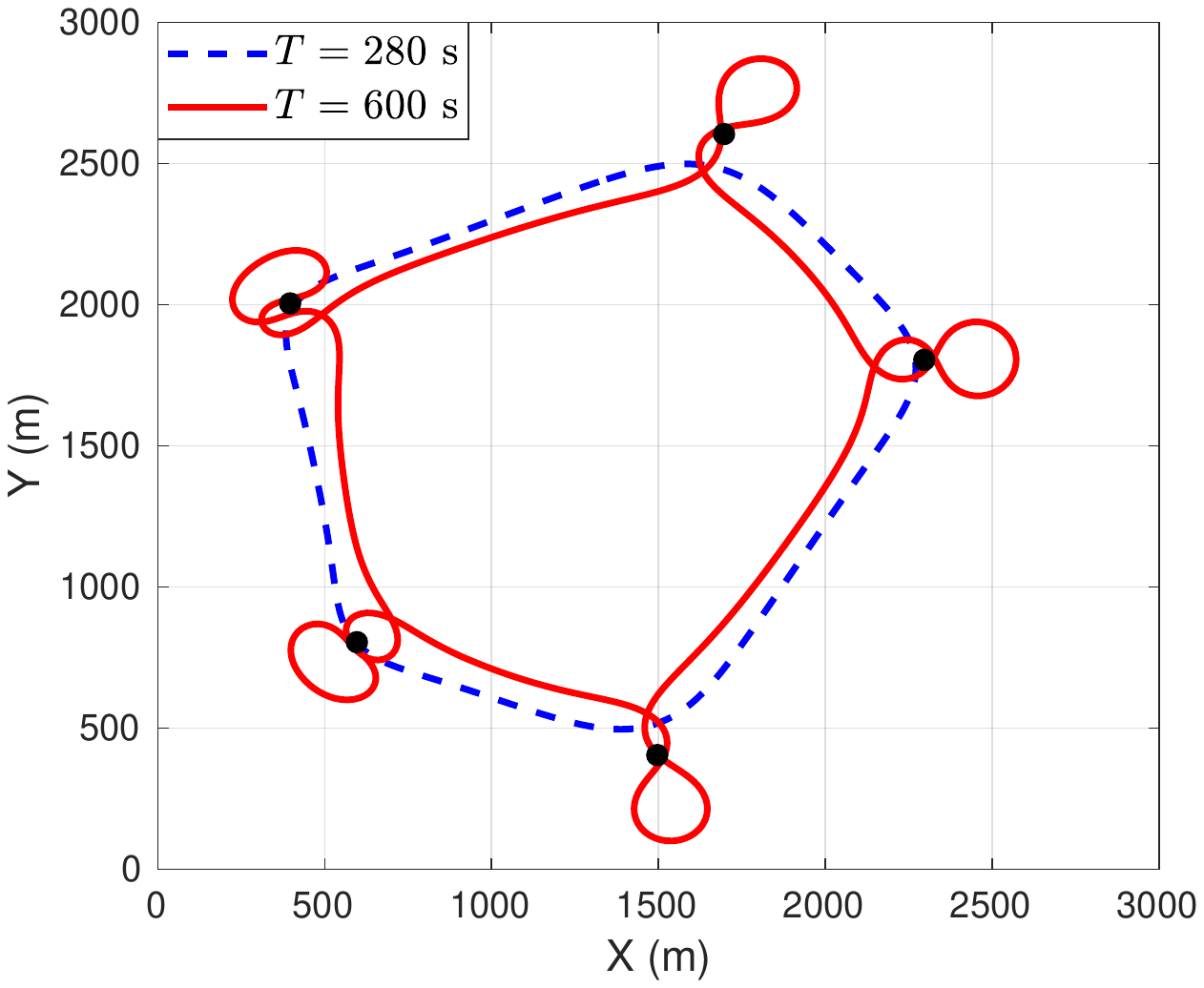}
    \vspace{-3.7cm}
    \caption{\label{p_traj}Trajectories obtained by the conventional time-discretization method. }
        \vspace{-1.5cm}
\end{minipage}%
    \hfill%
\begin{minipage}[t]{0.45\linewidth}
\vspace{-3.4cm}
    \hspace*{-2.1cm} \includegraphics[width=2\linewidth]{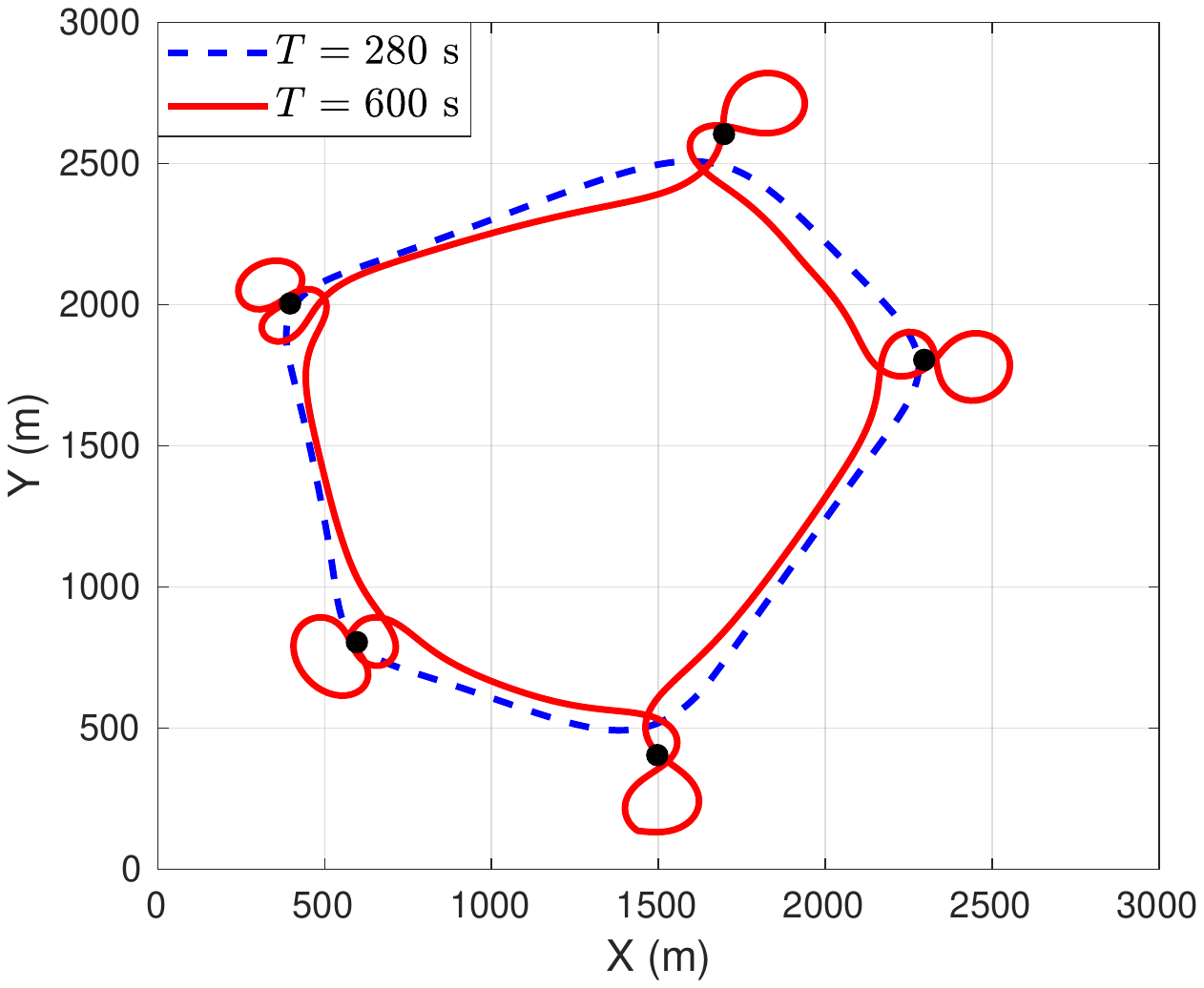}
    \vspace{-3.7cm}
    \caption{\label{r_traj}Trajectories obtained by the proposed RHO-based\\ method with $T_e=$80 s and $\overline{T}=$ 120 s.}
    \vspace{-0.2cm}
\end{minipage}
\end{figure}

We consider a system with $L$$=$$5$ GNs, which are randomly distributed in a square area of side length 3000 m. The following results are based on one realization of GNs' locations as shown in Fig. \ref{p_traj}. Note that the parameters related to the aircraft are set as $c_1$$=$$0.03125$, $c_2$$=$$1500$ such that the maximum endurance speed, i.e., the speed of the minimum power consumption, is 20 m/s and the corresponding propulsion power consumption is 1000 W. The altitude of the UAV is fixed at $H$$=$$100$ m, with the maximum UAV speed $V_{\mathrm{max}}$$=$$30$ m/s, minimum UAV speed $V_{\mathrm{min}}$$=$$5$ m/s, and maximum UAV acceleration $a_{\mathrm{max}}$$=$$3$ m/s$^2$. The transmit power of all GNs are set as $P_l$$=$$0.01$ W, $\forall l \in \mathcal{L}$ and the total available bandwidth is $B$$=$$1$ MHz. The reference channel power at a reference distance $d_0$$=$1 m is $\beta_0$$=$$-40$ dB with the noise power $\sigma^2$$=$$-169$ dBm. The path loss exponent is $\tilde{\alpha}$$=$$2$. Besides, the maximum segment lengths are set as $\Delta_1$$=$$30$ m and $\Delta_2$$=$$120$ m such that $\delta_1$$=$$1$ s and $\delta_2$$=$$4$ s, respectively. All simulations are run in MATLAB 2015b, which operates on Windows 10 with Intel-i5 3.2Hz PC and 8GB RAM.

\begin{figure}[htbp]
\begin{minipage}[t]{0.45\linewidth}
\vspace{-3.3cm}
    \hspace*{-1.8cm}\includegraphics[width=2\linewidth]{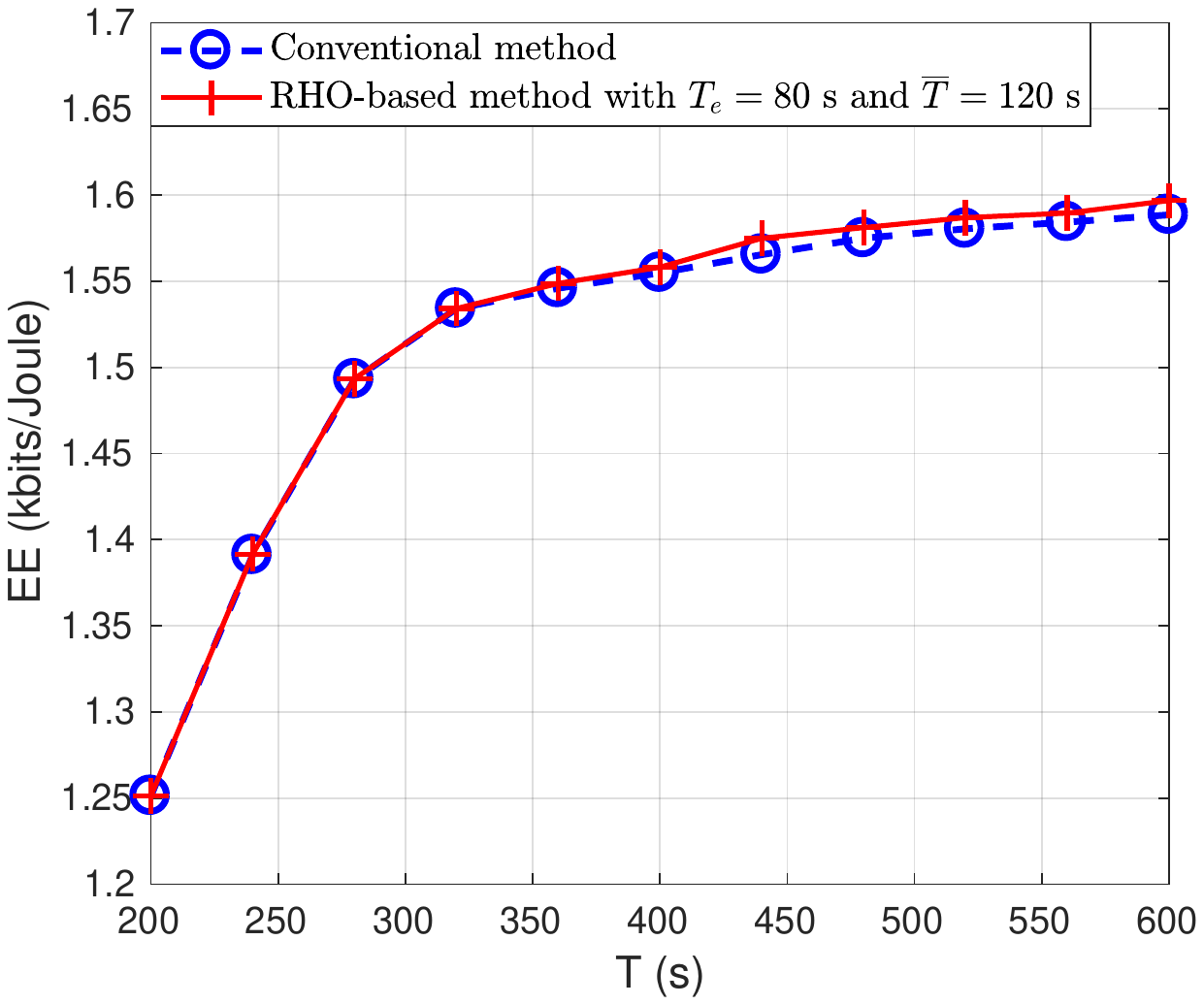}
     \vspace{-3.7cm}
    \caption{\label{ee}EE comparison.  }
      \vspace{-1cm}
\end{minipage}%
    \hfill%
\begin{minipage}[t]{0.45\linewidth}
\label{time}
\vspace{-3.3cm}
    \hspace*{-2cm} \includegraphics[width=2\linewidth]{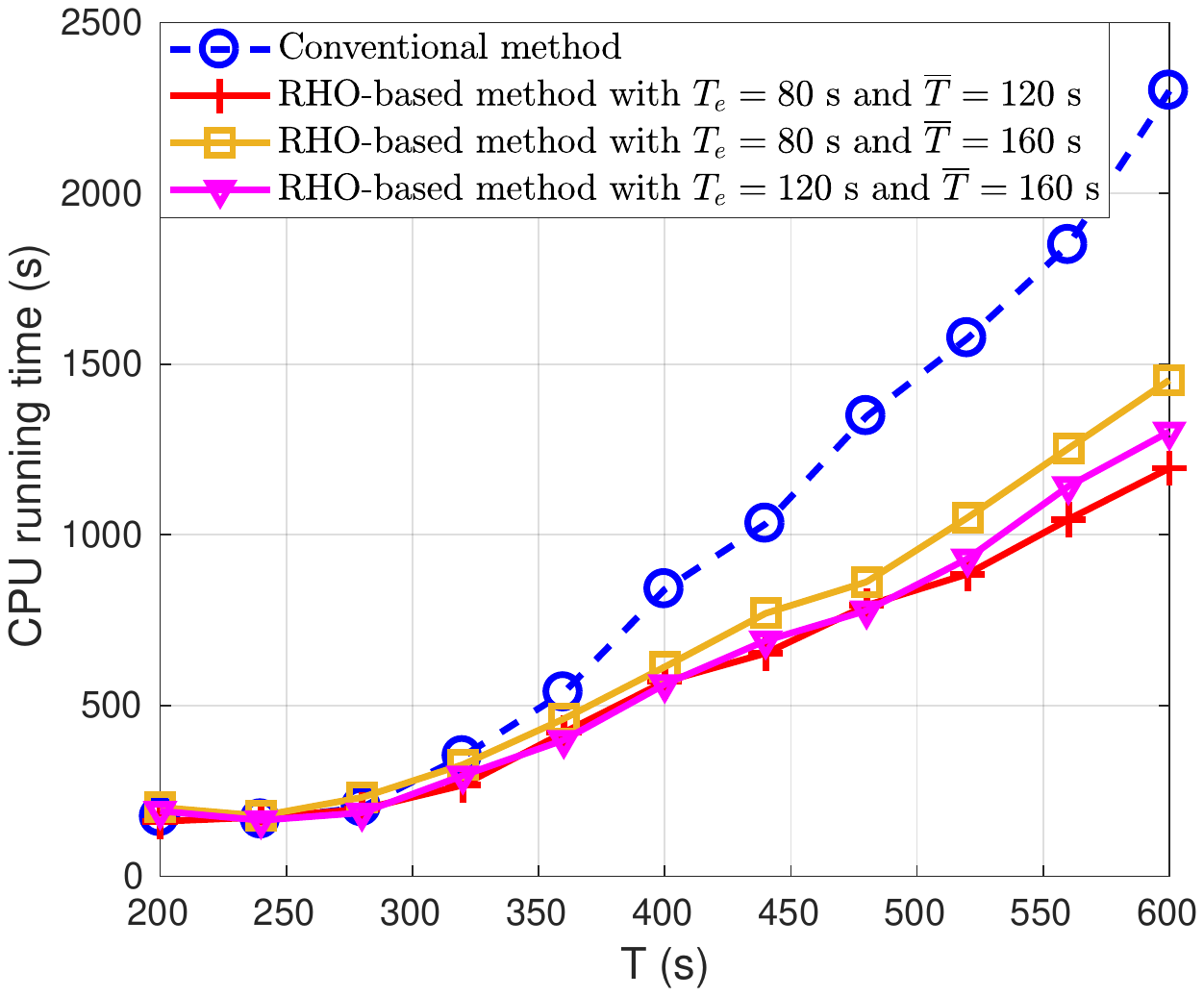}
     \vspace{-3.7cm}
\caption*{\!\!\!\!\!\!\!\!\!\!\!\!\!\!\!\!Fig. 5: Computation time comparison.\!\!\!\!\!\!\!\!\!\!\!\!\!\!\!\!\!\!\!\!~~}
     \vspace{-0.6cm}
\end{minipage}
\end{figure}

The trajectories obtained by the conventional method and the proposed RHO-based method are shown in Fig. \ref{p_traj} and~Fig.~\ref{r_traj}, respectively. It is observed that both methods lead to similar trajectories after convergence. In particular, when the given period is small, i.e., $T$$=$$280$ s, the UAV will approach~each GN to maximize the throughput, whereas when $T$ increases to $T$$=$$600$ s, the UAV follows an ``8" shape path above each GN, which is energy-efficient for fixed-wing UAVs while maintaining good communication channel with the corresponding GN being served \cite{zeng2017energy}. Fig. \ref{ee} compares the EE of the two methods, which demonstrates that the proposed RHO method achieves comparable EE with the conventional method, which is consistent with the similar trajectories shown in Figs. \ref{p_traj}~and~\ref{r_traj}.



The computation time of different methods is illustrated in Fig. 5. It is noted that when $T$ is relatively small, both the conventional time discretization and the RHO have similar computation time. This is expected since with a small $T$ given, the number of time-windows is also small, as such the complexity of solving the problem cannot be significantly reduced by the RHO-based method. However, as $T$ increases, the RHO-based method significantly outperforms the conventional method, due to the significantly reduced variables for solving the problem in each time-window.

Moreover, it is observed that under the same execution time $T_e=80$ s, the RHO-based method with a smaller time-window $\overline{T}=120$ s needs less computation time compared to that of $\overline{T}=160$ s. The reason is that under the same value of $T_e$ and with an approximately equal number of time-windows, a smaller value of $\overline{T}$ results in fewer variables in each time-window and thus less computation time. On the other hand, when the duration of the time-window $\overline{T}$ is fixed, since a smaller value of $T_e$ results in a larger number of total time-windows, it is observed that the computation time with smaller $T_e$ leads to slightly higher computation time. Therefore, the values of $T_e$ and $\overline{T}$ need to be carefully chosen for the RHO-based method to maximally reduce the computation complexity while not affecting the communication performance.

%

\section{Conclusion}

This letter proposes an RHO-based method for the joint optimization problem of UAV trajectory and communication resource allocation. By discretizing the problem with different accuracies, the number of variables for each time-window optimization is significantly reduced, thus leading to an overall lower computation complexity compared to the conventional method with uniform time discretization. The proposed method can be similarly applied to other setups such as path discretization, rotary-wing UAVs, and online trajectory design, etc., which will be left for our future work.


\ifCLASSOPTIONcaptionsoff
  \newpage
\fi

\bibliographystyle{IEEEtran}
\bibliography{refer}

\end{document}